\documentclass[10pt,letterpaper,twocolumn]{article}

\usepackage{ol2}
\usepackage[draft]{hyperref}
\usepackage{amsmath}
\usepackage{amssymb}
\usepackage{color}

\DeclareGraphicsExtensions{.jpg .png,.eps}

\begin{document}

\twocolumn[

\title{Nonlinear localized modes in Glauber-Fock photonic lattices}

\author{Alejandro J. Mart\'inez$^{1,2}$, Uta Naether$^{1,2}$, Alexander Szameit$^{3}$, and Rodrigo A. Vicencio$^{1,2}$}

\address{
$^1$Departamento de F\'isica and MSI-Nucleus on Advanced Optics, Facultad de Ciencias, Universidad de Chile, Santiago, Chile
\\
$^2$Center for Optics and Photonics (CEFOP), Casilla 4016, Concepti\'on, Chile \\
$^3$Institute of Applied Physics, Friedrich-Schiller-Universit\"at Jena, Max-Wien-Platz 1, 07743 Jena, Germany
}

\begin{abstract}
We study a nonlinear Glauber-Fock lattice and the conditions for the excitation of localized structures. We investigate the particular linear
properties of these lattices, including linear localized modes. We investigate numerically nonlinear modes centered in each site of the lattice. We
found a strong disagreement of the general tendency between the stationary and the dynamical excitation thresholds, and we give a new definition
based on both considerations.

\end{abstract}

 ]

\maketitle

\noindent Photonic crystals and waveguide arrays are one of the most fruitful areas of research in optics and nonlinear physics. During the last
years these ideal experimental and theoretical scenarios enabled the prediction and demonstration of many fundamental properties of the general area
of discrete nonlinear systems~\cite{rep1,rep2}. Studies of periodic and non-periodic systems allowed to formulate the necessary conditions for the
excitation of nonlinear localized structures (i.e., \textit{discrete solitons}) in different dimensions and geometries. For example, surface
solitons have been an intense area of research~\cite{surface} where stationary localized solutions exist only above a power threshold.

The development of new experimental techniques has allowed a variety of interesting configurations. In particular, we mention here
the laser direct-writing approach using focused pulsed femtosecond laser radiation~\cite{arrays}. The possibilities range from
ordered to disordered systems in 1D and 2D settings. For example, main properties of chirped arrays have been theoretically~\cite{olchirped1} and
experimentally~\cite{olchirped2} analyzed. Very recently, in the context of quantum optics in waveguide arrays, Glauber-Fock (GF) photonic
lattices~\cite{ol1,prl1} have been studied. Authors found, for example, the appearance of a new family of quantum correlations which are absent in
uniform arrays.

In this letter, we study the excitation of nonlinear localized states in a GF lattice. This system corresponds to a single-mode optical waveguide
array whose coupling between sites increases as a square root. In the coupled-mode approximation, the evolution of the field amplitude is described
by a discrete nonlinear Schr\"odinger-like equation~\cite{rep1,rep2,ol1,prl1}:
%
\begin{equation}
-i\frac{d E_n}{d z}=V_{n+1}E_{n+1}+V_nE_{n-1}+\gamma|E_n|^2E_n\ ,
\label{dnls}
\end{equation}
%
where $E_n$ denotes the amplitude of the mode in the $n$-th waveguide for an array of $N$ sites. $z$ corresponds to the propagation coordinate along
the waveguide, $\gamma$ to the nonlinear parameter and $V_n=\sqrt{n} V_1$ to an increasing coupling coefficient (GF lattice). Without loss of
generality, we define $V_1=1$ and $\gamma=1$ (focusing case). To characterize the solutions, we use the integral power (P) and the
participation ratio (R) defined as
\begin{equation}
P\equiv\sum_{n=1}^N|E_n|^2\ \ \ \ \ \text{and}\ \ \ \ \ R\equiv\frac{P^2}{\sum_{n=1}^{N}|E_n|^4}.
\end{equation}
%

%
\begin{figure}[h]
\centering
\includegraphics[width=8.3cm]{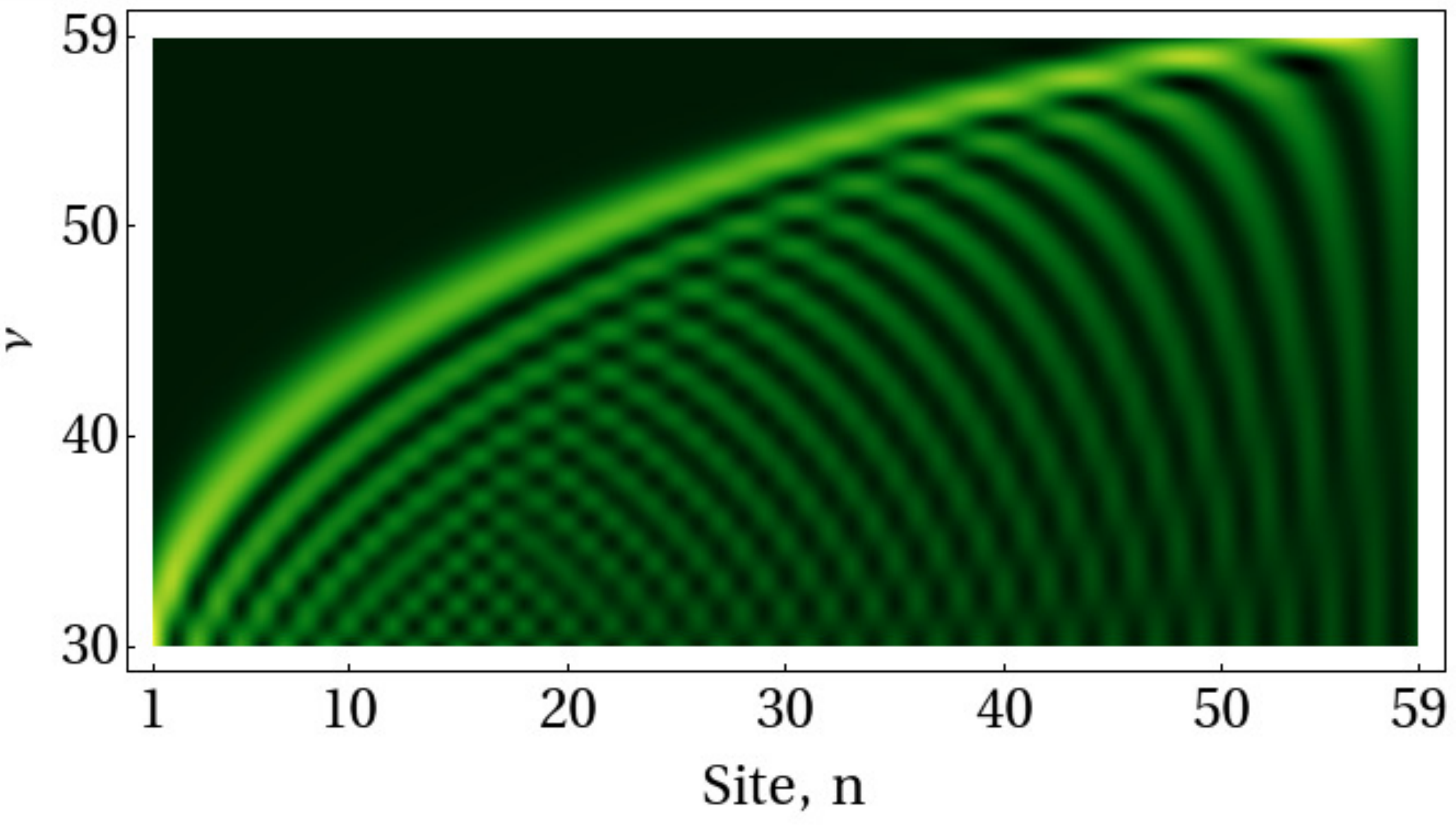}
\caption{Intensity distribution ($|E_n^{\nu}|^2$) of linear modes.}
\label{fig1}
\end{figure}
%
Linear modes of model (\ref{dnls}) are given by
$E_n^{\nu}(z)=\sqrt{n!} 2^{-n/2}
H_n(\lambda_{\nu}/\sqrt{2})\exp(i\lambda_{\nu} z)$, with $H_n$ being an Hermite polynomial~\cite{ol1}. The eigenvalue spectrum $\{\lambda_{\nu}\}$ does not form a compact band, but rather a set of well separated discrete frequency levels. The system is invariant under the staggered-unstaggered transformation: 
\[
E_n^{[-\nu+(N+1)/2]}=(-1)^n E_n^{[\nu+(N+1)/2]}\ \ \forall \nu \in
\{1,(N-1)/2\},
\]
with $\lambda_{[\nu+(N+1)/2]}=-\lambda_{[-\nu+(N+1)/2]}$. Therefore, only half of the modes are shown in Fig.\ref{fig1} for $N=59$, following Ref.~\cite{prl1}. The spatial distribution of the linear modes is very inhomogeneous. For modes with frequencies close to zero ($\nu\sim 30$), light is extended across the array with a well defined main lobe centered at different positions of the lattice [see Fig.\ref{fig1}]. For modes with frequencies closer to the band edges ($\nu\sim 1$ or $59$), profiles are well localized. [For an homogenous DNLS lattice, all linear modes are spatially extended ($R\sim 2N/3$) without any main lobe or peak~\cite{rep1,rep2}].

As a consequence of the linear properties (mode profiles and frequency gaps), nonlinear modes will exist in different frequency domains. For example,
modes $\nu =1$ and $59$ are the most localized ones ($R\approx 9$), both having a peak located around $n\sim 55$. These modes have no extended tail
and they are a result of the ramping in the coupling coefficient (for chirped systems, there is always at least one linear localized mode close to
one of the borders of the lattice~\cite{olchirped1,olchirped2}). Therefore, we expect to find a nonlinear solution bifurcating with zero power
threshold close to this region. On the other hand, the center mode ($\lambda_{30}=0$, $R\approx 18$) has a main peak located at $n=1$. This mode
possesses a non-negligible tail which will forbid the excitation of a zero-threshold fundamental nonlinear solution (one-peaked bright solitons, i.e.
an \textit{odd mode}~\cite{rep2}). In addition, modes $\nu= 28$, $29$, $31$ and $32$ also possess a main peak located at $n=1$. Therefore, we expect
that different linear modes - at different frequencies - contribute in the formation of the profile/tail structure while the solution decreases its
power and interact with the linear spectrum. For example, modes $\nu= 26$, $27$, $33$ and $34$ possess a main peak located at $n=2$. Therefore, the
respective nonlinear mode will not have an unique frequency origin, and it will probably experience some power threshold. By observing
Fig.~\ref{fig1}, we expect that the power and frequency thresholds increase as the position of the linear mode peak increases with $n$ (up to the
region where we expect a zero threshold). Then, the power threshold should increase again because there is no further linear mode peaked at
$n>56$.

We look for stationary solutions of Eq.(\ref{dnls}) of the form $E_n(z)=A_n\exp(i\lambda z)$, where $A_n$ are real numbers and $\lambda$ corresponds to the solution frequency. We compute one-peak localized stationary solutions by using a Newton-Raphson iterative method starting at $\lambda\gg \lambda_{59}\approx 14$. Once we get a numerical solution we make a sweep in $\lambda$ to construct the whole family. For each solution, we perform a standard stability analysis~\cite{surface} and, as a general convention, we use black (gray) lines for stable (unstable) solutions.
%
\begin{figure}[h!]
\centering
\includegraphics[width=8.5cm]{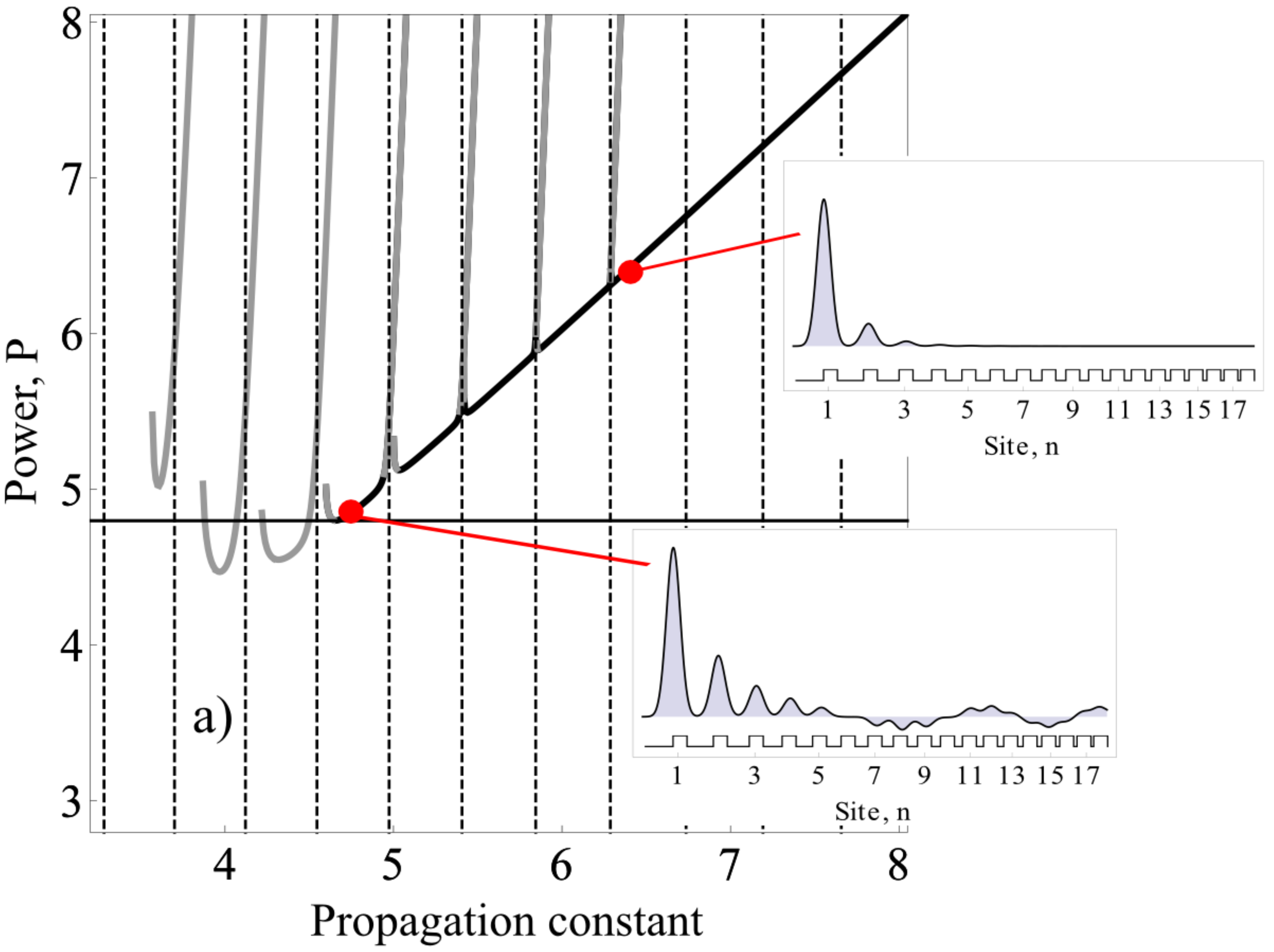}
\includegraphics[width=8.5cm]{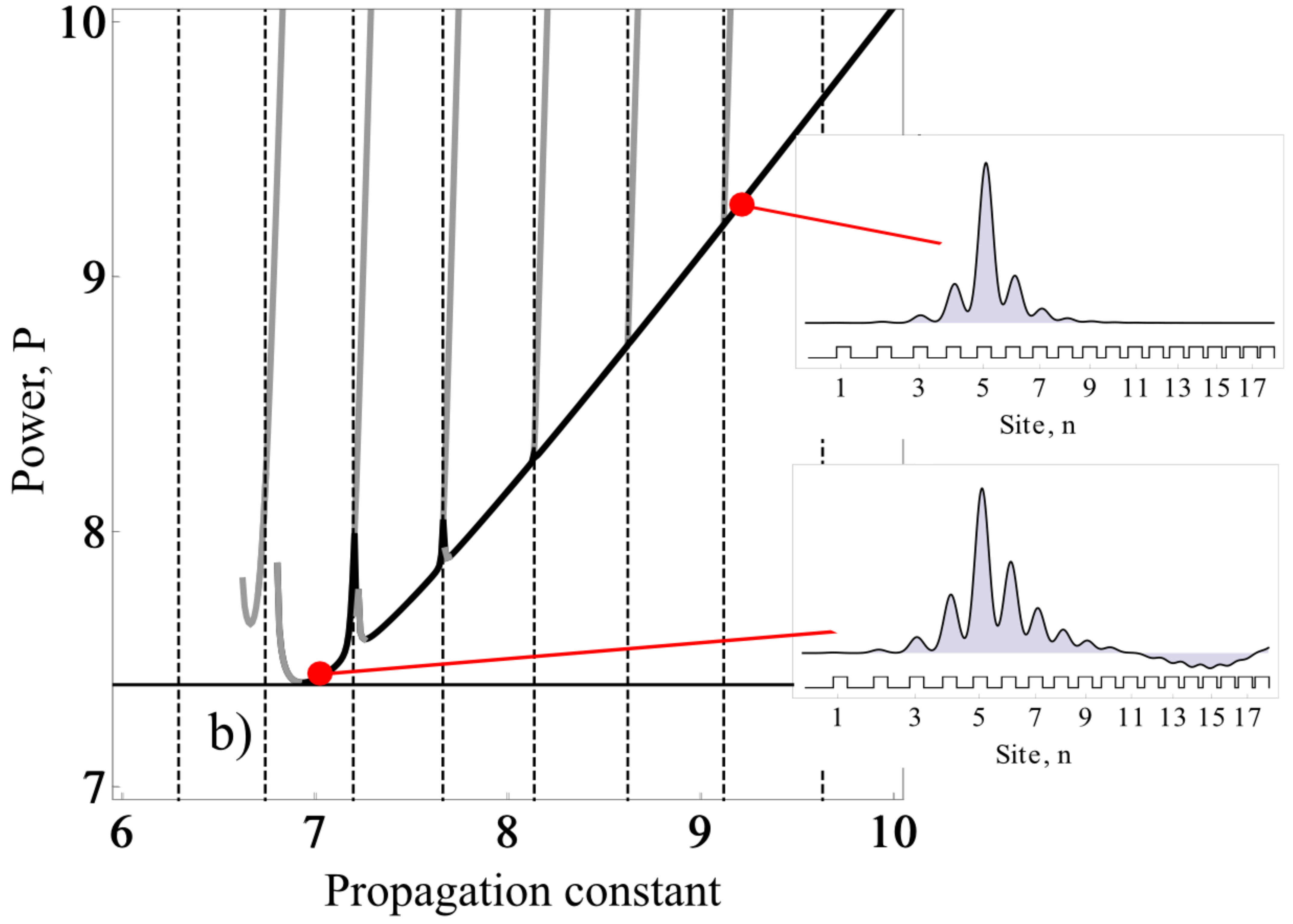}
\includegraphics[width=8.cm]{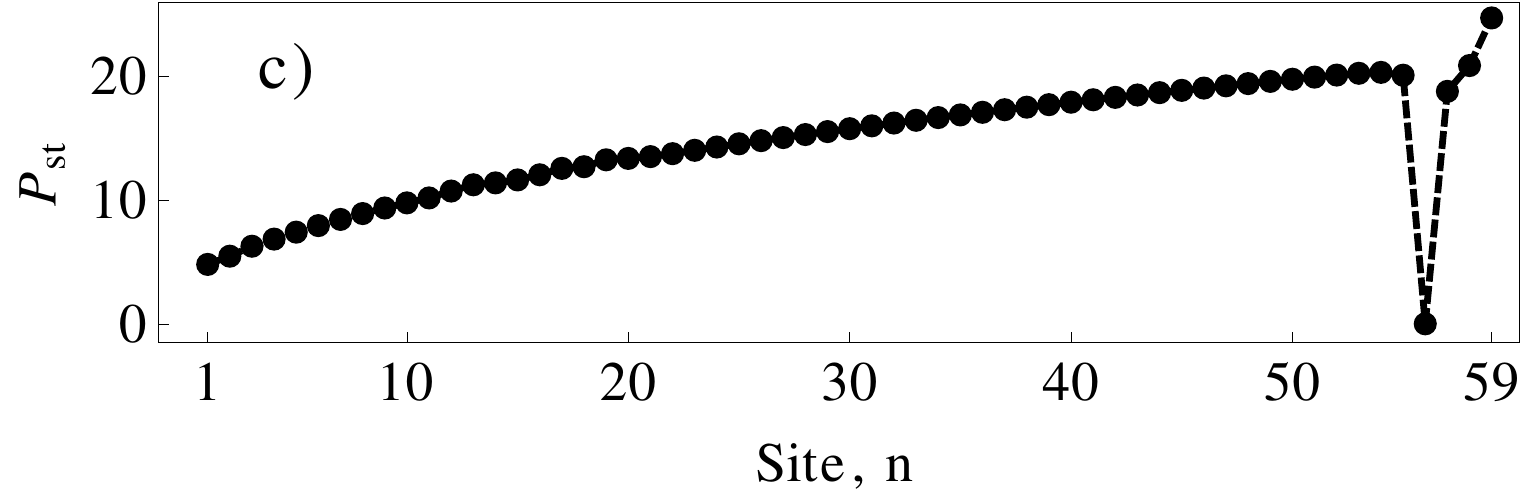}
\caption{(a) and (b) $P$ versus propagation constant ($\lambda$) for solutions peaked at site $n=1$
and $n=5$, respectively (vertical dashed lines correspond to different $\lambda_{\nu}$).
Insets show profiles $\{A_n\}$ for parameters indicated in each figure.
(c) $P_{st}$ versus the peak position $n$.}
\label{fig2}
\end{figure}
%
Fig.\ref{fig2}(a) shows a Power versus Frequency diagram for solutions with a main peak located at $n=1$. Depending on the accuracy of our numerical continuation, we detect some ``resonances'' of the nonlinear solutions with some linear band modes (vertical dashed lines indicate different
$\lambda_{\nu}$). If we numerically continue these solutions - by increasing the frequency - we see how the tails increase with an structure similar
to the linear modes at those regions [see divergent curves in Figs.\ref{fig2}(a)-(b)]. Solutions peaked close to $n=1$ present similar features; see,
for example, Fig.\ref{fig2}(b). Insets in Figs.\ref{fig2}(a)-(b) show some profiles for these families. We see how high frequency (power) solutions
are well localized and how low frequency solutions interact with the linear modes (see the tails of these profiles). If we excite solutions peaked at
larger $n$, we find a phenomenology typical for surface modes~\cite{surface}; i.e., the power and frequency decrease up to a point where
$\partial P/\partial \lambda<0$, solutions become unstable and are not allowed to exist below a power threshold. In contrast, the solution
peaked at $n=56$ exhibits a power that drops to almost zero, because it bifurcates from the linear localized mode $\nu=59$. We characterize
all the stationary solutions by defining the \emph{stationary power threshold ($P_{st}$)} as the last set $\{\lambda,P\}$ before the solution becomes
unstable [horizontal lines in Figs.\ref{fig2}(a)-(b)]. Fig.\ref{fig2}(c) shows $P_{st}$ versus the site where the solution has its main peak. We see
an approximately square-root-dependence of the power increment from $n=1$ up to the region $n\sim56$, where this threshold abruptly
drops to zero. For the further solutions, the stationary threshold increases again up to a maximum.

In a next step we want to determine the dynamical power threshold for the excitation of one-site solutions centered in different sites of
the lattice. We integrate numerically model (\ref{dnls}) with a one-site input condition $E_n(0)=\sqrt{P_0} \delta_{n,n_0}$, where $P_0$ is the input
power and $n_0$ the input position. For the set $\{n_0,P_0\}$ we compute the space-averaged fraction of power: $T\equiv 1/(P_0 z_{max})
\int_0^{z_{max}}|E_{n_0}(z)|^2dz$, remaining at the initial waveguide after a fixed propagation distance $z_{max}$.
%
\begin{figure}[h!]
\centering
\includegraphics[width=8.cm]{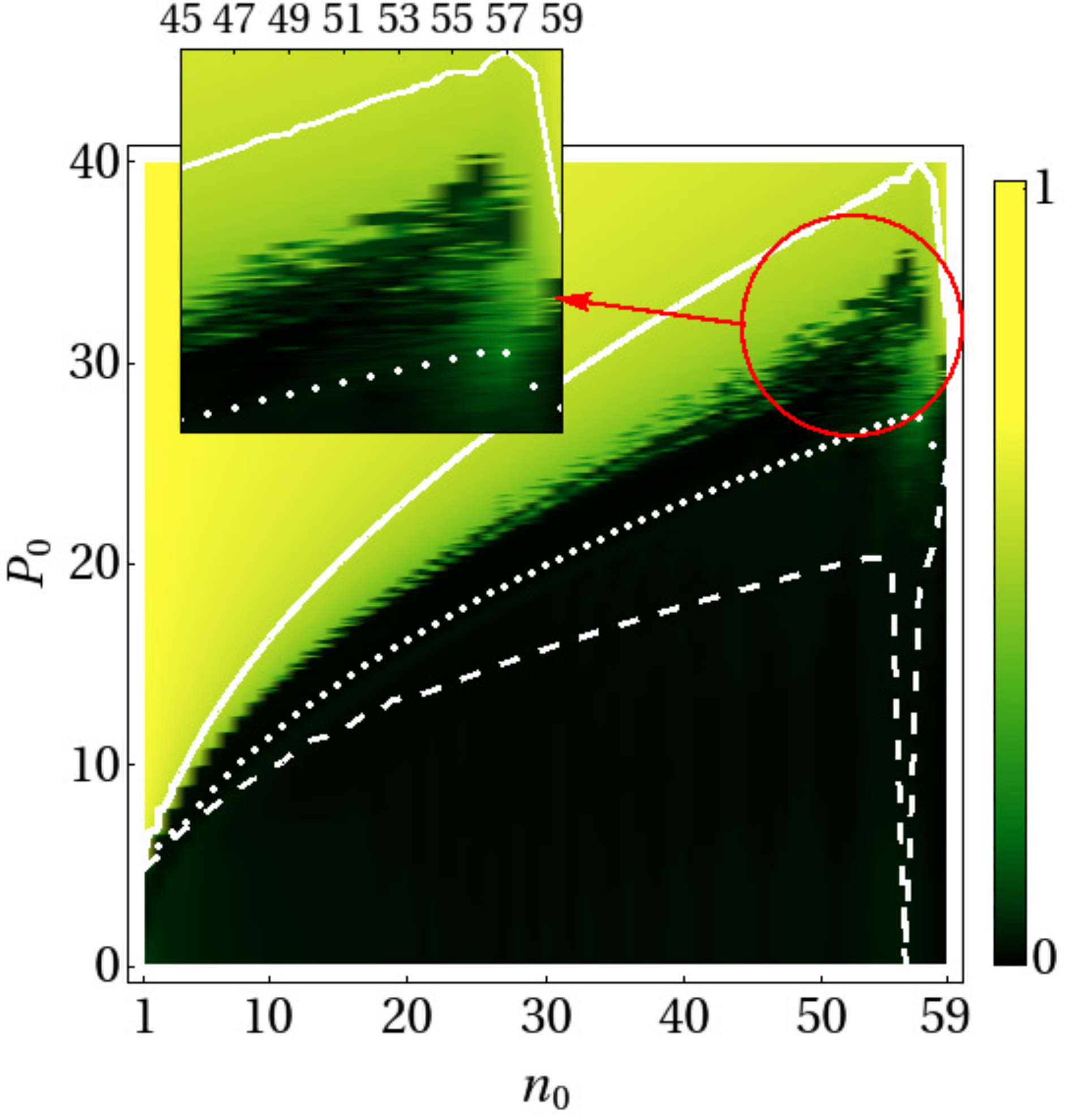}
\caption{$T$ as a function of $P_0$ and $n_0$. Continuous, dashed and dotted white lines correspond to $T=0.8$, $P_{st}$ and $P_{ps}$, respectively.
Inset: zoom of region $n\in \{45,59\}$.} \label{fig3}
\end{figure}
%
Fig.\ref{fig3} shows our results where $T=1$ (lighter color) means that $100\%$ of the light is confined in the input waveguide whereas
$T=0$ (darker color) means that all the light has escaped from the input position. In this figure, we have included a continuous white line for
$T=0.8$ as an arbitrarily chosen localization indicator. This line shows an increasing dynamical threshold for nonlinear states with an
increasing position. Then, this threshold just decreases up to some minima at the last site of the array.  A white dashed line indicates the
stationary power threshold $P_{st}$. The main tendency is corroborated and the dynamical threshold is always larger than the stationary one. However, in the region of zero stationary power threshold there is an important disagreement. For this reason, we introduce the \emph{pseudo-stationary power threshold} ($P_{ps}$). We define $P_{ps}$ as the value of the stationary power $P$ for which the peak amplitude is an $80\%$ of the total power ($|A_{n_0}|^2=0.8P$). This quantity is computed for stationary solutions and is represented with a white dotted line in Fig.~\ref{fig3}.
Our definition incorporates the experimental restriction of one-site input excitations for exciting highly localized states ($P_{ps}$
implies that most of the power is contained at the peak). The experimental excitation of stable fundamental solutions like the ones sketched in
insets of Figs.\ref{fig2}(a)-(b) is not a trivial issue due to the long tail of the solutions. From Fig.\ref{fig3} we see how the tendency of our definition agrees perfectly with the dynamical result. Nevertheless, the pseudo-stationary prediction will always be smaller than the dynamical one (for example, bulk DNLS solutions have a stationary threshold equal to zero and a dynamical one around $4$).

Additionally, we have studied systems with different functional forms of $V_n$ (exponential, quadratic, linear and logarithmic). All these
systems possess a similar band structure in the sense of non-compactness and spatial distribution of linear modes. The numerical computation of $T$
shows similar features than the ones sketched in Fig.\ref{fig3}, except the global shape of the increment which is proportional to the shape of the
particular $V_n$ function. In order to experimentally prove the threshold behavior and observe nonlinear localized modes in such structures,
the range of the coupling function $V_n$ is constrained to the length of the sample. A large value of $V_n$ implies a strong coupling and a faster
transport from one guide to another and, therefore, requires a very long crystal for observing such phenomenology.

In conclusion, we studied the linear properties and the stationary and dynamical excitation of nonlinear localized modes in GF lattices. Contrary to
homogenous systems, we found that close to the lattice-edges solutions possess smaller stationary thresholds. We introduced a new quantity, the
$P_{ps}$, which showed its utility when comparing stationary and dynamical properties.

The authors acknowledge support from CONICYT fellowship, FONDECYT Grant 1110142, Programa ICM P10-030-F, Programa de Financiamiento Basal de CONICYT (FB0824/2008) and the German Ministry of Education and Research (Center for Innovation Competence program, grant 03Z1HN31).

\end{document}